# Semi-classical modelling of second-order spontaneous down-conversion measured over 10 decades of pump intensity in a type II phase-matched KTP crystal

Julien Bertrand, Véronique Boutou, and Benoît Boulanger*

Université Grenoble Alpes, CNRS, Grenoble INP, Institut Néel, 38000 Grenoble, France
*benoit.boulanger@neel.cnrs.fr

**Abstract**

We performed a second-order spontaneous parametric down-conversion (SPDC) experiment in a Type II phase-matched $KTiOPO_4$ crystal pumped at 532 nm giving birth to a signal beam at 1037 nm and an idler one at 1092 nm along the x-axis of the crystal. Three pump sources have been used: in the continuous-wave, nanosecond (200 ns, 1 kHz) and picosecond (15 ps, 10 Hz) regimes. That allowed us to cover a pump intensity range from 2.4 W.cm$^{-2}$ to 3.7 GW.cm$^{-2}$ which led to the generation of twin photons with a flux from $1.1 \times 10^4$ Hz to $1.2 \times 10^{21}$ Hz, respectively, the corresponding quantum efficiencies being of $1.8 \times 10^{-11}$ and $9.2 \times 10^{-4}$. We identified two SPDC regimes regarding the pump intensity: one below 30 MW.cm$^{-2}$ for which the flux increases linearly, and the other one above 30 MW.cm$^{-2}$ for which the flux behaves exponentially. We proposed a semi-classical model under the undepleted pump approximation based on both the quantum fluctuations of vacuum for the seeding of the process and a propagation step ensured by classical fields, which allowed a satisfying description of these two regimes, with a much better agreement at high intensity than the quantum models based on the nonlinear momentum or the nonlinear Hamiltonian operators.

## 1. Introduction

Twin photons have profoundly influenced the history of nonlinear and quantum optics through their wide range of applications and the paradigmatic place they occupy in the generation of new quantum states of light. Twin photon generation, also known as spontaneous parametric down-conversion (SPDC), is a second-order nonlinear process in which a pump photon is split into two photons of lower energy. Nowadays, bulk crystals, photonic crystals or waveguides are used as nonlinear media.[1] Photons are emitted in pairs, which radically alters their behavior, at the very heart of several non-classical properties, such as for example the squeezing of quantum fluctuations, or quantum entanglement that is the most intriguing.[2, 3, 4] These quantum aspects are the key element in a large number of demonstrations such as quantum cryptography, quantum teleportation and quantum information in the broadest sense, or the detection of gravitational waves.[5, 6, 7, 8] While the quantum properties of photon pairs are extremely well known, it is still difficult to model their generation in order to know the flux as a function of pump intensity. Major efforts have been made in recent years to develop quantum models, the most successful in our opinion being those based on the nonlinear momentum operator and the nonlinear Hamiltonian operator.[9,10, 11] The present work falls within this framework, with the ambition of further improving predictivity. We have achieved this by proposing and developing a semi-classical model that provides a satisfactory description of the generation of photon pairs over 10 decades of pump intensity, *i.e.* from 1 W.cm$^{-2}$ to 10 GW/cm$^{-2}$. We then set up an SPDC experiment based on three sources operating in the CW, nanosecond and picosecond regimes, with a type II phase-matched bulk KTP crystal as the nonlinear medium.

## 2. Measurements

2.1. Experimental design

We aim to quantitatively and accurately measure the twin photon flux of SPDC over a wide pump intensity range of several decades. For that purpose, we design an experiment which uses a collinear type II phase-matched 1-cm-long x-cut KTP crystal pumped at 532 nm. The energy and momentum conservations write:

$$\hbar\omega_p - \hbar\omega_s - \hbar\omega_i = 0 \tag{1}$$

and

$$\Delta k(\omega_p, \omega_s) = \frac{\omega_p}{c} n_p - \frac{\omega_s}{c} n_s - \frac{\omega_i}{c} n_i = 0, \tag{2}$$

with $n_p = n_y(\omega_p)$, $n_s = n_z(\omega_s)$  $n_i = n_y(\omega_i = \omega_p - \omega_s)$, *y* and *z* referring to the dielectric frame (*x,y,z*) of KTP. The indices "p", "s" and "i" refer to the pump, signal and idler, respectively.

In this configuration, the pump field and the down-converted fields are coupled by the nonlinear coefficient $\chi^{(2)}_{24}(532\ nm) = 5.3\ pm.V^{-1}$ of KTP.[12] Moreover, using the Sellmeier equations, it comes for the phase-matching wavelengths of the twins : $(\lambda_s = 1037\ nm, \lambda_i = 1092\ nm)$.[13]

**Figure 1** details the experimental setup that we used for the quantitative measurement of SPDC over a large pump intensity range of 10 decades.

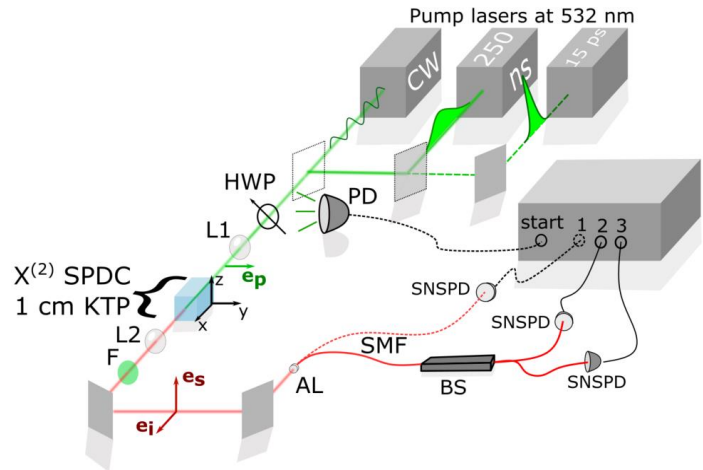

**Figure 1:** Experimental setup used for performing CW, nanosecond and picosecond Type II SPDC in a x-cut KTP crystal pumped at 532 nm. (**e$_p$**, **e$_s$**, **e$_i$**) stand for the polarizations of the pump, signal and idler beams, respectively.

Three different sources at 532 nm were used interchangeably: a continuous-wavelength (CW) laser (Spectra Physics Millenia eV), a 1 kHz nanosecond (ns) laser (Coherent Evolution 30, $\tau = 200$ ns) and a 10 Hz picosecond (ps) source ($\tau = 15$ ps). **Table 1** summarizes the experimental parameters for each experimental setup. A half-wave plate (HWP) allowed to polarize the pump beam along the y-axis of KTP. The pump beam was focused using an optical lens (L1) into a 1-cm-long x-cut KTP. L1 was chosen so that the Rayleigh length of the pump beam, deduced from the beam waist *radius*, is greater than the length of the crystal. By this way, the SPDC experiment was performed with parallel beams in order to minimize the "degree" of non-collinearity. The generated beams were further collimated with an optical lens (L2), filtered with a filter-set (F) including notch filters at 532 nm to reject the pump and injected into a single mode fiber (SMF) by use of an aspheric lens (AL). The focal length $f_2$ of lens $L2$ was always chosen so that the pump beam size fits well the aspherical lens diameter to maximize the injection efficiency of the SPDC signal in the SMF.

**Table 1:** Experimental parameters for each setup.

| Pump source | CW | ns | ps |
|---|---|---|---|
| Pulse duration ($\tau$) | $NA$ | $200\ ns$ | $15\ ps$ |
| Repetition rate ($f_{rep}$) | $NA$ | $1 kHz$ | $10\ Hz$ |
| Pump intensity range ($I_p$) [$W \cdot cm^{-2}$] | $[1 - 7.10^3]$ | $[7.10^3 - 10^5]$ | $[160.10^6 - 6.10^9]$ |
| Focus lens ($f_1$) [mm] | 500 | 500 | 750 |
| Rayleigh length $2 \times Z_0$ [cm] | 7.2 | 70 | 9.4 |
| Waist *radius* $W_0$ [$\mu m$] | 78 | 250 | 90 |
| Collection lens ($f_2$) [mm] | 300 | 300 | 400 |
| Effective numerical aperture ENA for SPDC collection | 0.041 | 0.041 | 0.03 |
| Detection setup | Coincidence Counting 2 SNSPD | Coincidence Counting 2 SNSPD | Triggered Detection 1 PD/1SNSPD |

Highly sensitive MoSi Supraconducting Nanowire Single Photon Detectors (SNSPDs) from IDQuantique were used as detectors. The configuration of the detection stage depends on the pump source. For the CW and kHz ns lasers (solid lines in Figure 1), the signal was separated on a balanced Beam Splitter (BS) and coincidence measurements were performed on channel 2 and 3. In the case of the ps pump laser, the detection on channel 1 was triggered by the signal of a photodiode (PD) on a pump leak ; actually, because of the short time duration of the pulse, *i.e.* $\tau = 15\ ps$, the coincidence measurements were not possible so that a Silicon based photodiode DET 110 (Thorlabs) installed on a leak of the pump laser is used as external trigger (dotted line in Figure 1). The protocols of detection are described hereafter.

- For the CW experiment, the quantitative estimation of the twin photon rate as well as the transmission of the setup were performed by comparing the coincidence to single *ratio* on the SNSPD.[11,14] For the lowest pump intensity, the integration time was pushed up to 30 minutes. The upper limit for the pump intensity is linked to the available laser power.

- For the kHz nanosecond measurements, the coincidence experiment was performed in a particular hybrid time regime where the time duration of the pulse $\tau_{pulse} = 200\ ns$ was of the order of the SNSPDs recovery time $\tau_{recovery} = 30\ ns$ leading to an artificial lowering of the detection efficiency, in particular for pump intensity above $10^5\ W.cm^{-2}$. In the range of the measurements, the detection efficiency had to be calibrated to retrieve reliable quantitative measurements. This has been done on *data* points using the previous CW pump experiment results. The upper boundary in term of pump intensity was limited for this pump source by the nonlinear detection efficiency decay due to $\tau_{recovery} \approx \tau_{pulse}$.

- In the picosecond regime, the coincidence detection setup using two SNSPDs was inoperant. The twin photons beam signal was then detected by use of one SNSPD. In this high intensity pump regime, the twin photons energy was accurately calibrated thanks to free space measurements for the greatest energies using a pico-joulemeter molectron J3S10. The energy was strong enough that it had to be attenuated using neutral densities to avoid breakdown of the SNSPD. The densities were chosen so that the average signal photon number *per* pulse be $N_{signal} \ll 1$ in order to statistically avoid any detection problems induced by $\tau_{pulse} = 15\ ps \ll 30\ ns = \tau_{recovery}$. For this experiment, the signal-to-noise *ratio* limited the measurement: the noise was due to optical background photons losses from the pump laser that were then synchronized with the trigger.

## 2.2 Experimental results

At pump intensities higher than $1\ GW.cm^{-2}$, the twin photons energy was high enough to record a spectrum using a USB 2000 Ocean Optics spectrometer. Two typical spectra, taken with a pump intensity $I_p \approx 2.3\ GW.cm^{-2}$ are given on **Figure 2** together with the intensity beam profile of the twin photons (inset of Figure 2). They have been registered with or without an iris before the spectrometer, in order to diaphragm the profile. The first one (with the iris), in orange on Figure 2, corresponds to the center of the SPDC beam profile spotted by the orange circle in the inset: the signal and idler wavelengths are in good agreement with the theoretical expectation from Equations (1) and (2), *i.e.* $\lambda_s = 1037\ nm$ and $\lambda_i = 1092\ nm$ (vertical dashed lines). In contrast, the second spectrum (without the iris), in black, corresponding to the full SPDC beam profile, is broaden because of non-collinear contributions.

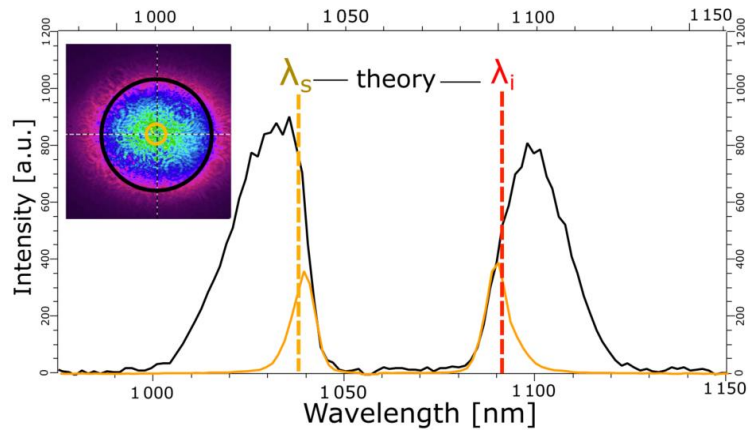

**Figure 2:** Twin photons spectrum. The dark line is the spectrum obtained when the complete emission cone is collected. The orange line corresponds to the spectrum obtained by diaphragming the non-collinear contributions. On the beam profile in inset, the numerical aperture of the collection lens is drawn as a circle in each case (black and orange circle). The vertical dashed lines correspond to the theoretical expectations.

On **Figure 3** are plotted the twin photons flux measured as a function of the pump intensity for the three runs: CW, ns and ps. Two regions are distinguishable: the first one, for $I_p$ below of about 30 $MW.cm^{-2}$, corresponds to a linear increase of the twin-photon flux; the second one, above 30 $MW.cm^{-2}$, relates to an exponential increase.

The experimental data are compared with the theoretical calculations further developed.

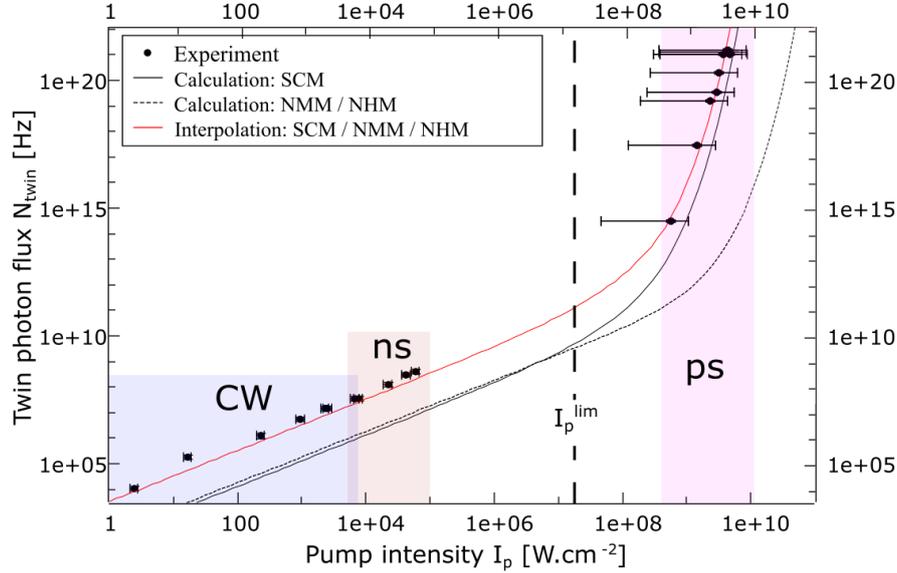

**Figure 3:** Measured and calculated SPDC photon flux generated in a 1-cm long x-cut KTP crystal as a function of the incident pump intensity. $I_p^{lim} \approx 30\ MW.cm^{-2}$ is the pump intensity delimiting the low and high pump intensity regimes of SPDC. SCM stands for Semi-Classical Model, NMM for Nonlinear Momentum Model and NHM for Nonlinear Hamiltonian Model.

### 3. Modelling

### 3.1 Quantum models

We first compared the experimental data of Figure 3 with two reliable quantum models that already exist: one based on the nonlinear momentum operator, and the other one on the nonlinear Hamiltonian operator.[10,11]

*3.1.1 The nonlinear momentum model*

The nonlinear momentum model (NMM) describes the quantum field in term of space dependent spectral mode operators $a(\omega, Z)$, and considers a single collinearly propagating spatial mode, assuming the undepleted pump approximation (UPA).[10,15] The corresponding calculation of the twin photon flux is detailed in **Appendix A**. Two regimes can be identified according to the parameter $C^{(2)}$ expressed as:

$$C^{(2)}(\omega) = 2\pi\, f^{(2)}(\omega)^2 I_p(0) {\chi_{eff}^{(2)}}^2 - \frac{\Delta k(\omega)^2}{4}, \tag{3}$$

with

$$f^{(2)}(\omega) = \sqrt{\frac{\omega(\omega_p - \omega)}{16\pi\epsilon_0 c^3 n_p(\omega_p) n_s(\omega) n_i(\omega_p - \omega)}} \qquad (4)$$

$I_p(0)$ is the pump intensity at the entrance of the nonlinear crystal, $\chi_{eff}^{(2)}$ is the effective coefficient that is equal to $\chi_{24}^{(2)}$ in the present case, and $\Delta k$ is given by Equation 2 where $n_p$, $n_s$ and $n_i$ are the refractive indices defined in section 2.1.

Then it comes for the signal photon flux:

$$N_s(\omega, Z) = \begin{cases} \dfrac{I_p(0)\chi_{eff}^{(2)^2} f^{(2)}(\omega)^2}{C^{(2)}(\omega)} \sinh^2\left(\sqrt{C^{(2)}(\omega)}Z\right) & \text{if } C^{(2)}(\omega) > 0 \\ I_p(0)\chi_{eff}^{(2)^2} f^{(2)}(\omega)^2 Z^2 \text{sinc}^2\left(\sqrt{|C^{(2)}(\omega)|}Z\right) & \text{if } C^{(2)}(\omega) < 0 \end{cases} \qquad (5)$$

Z is the space coordinate along the direction of propagation.

Finally, the twin photon flux writes:

$$N_{twin}(Z) = N_s(Z) = N_i(Z) = \int_0^{+\infty} d\omega\, N_s(\omega, Z), \qquad (6)$$

The corresponding curve is plot in Figure 3. It appears that the behaviour of the experimental data is well described, but there is a strong discrepancy in term of magnitude, from one order of magnitude lower to eight orders of magnitude, as shown in **Table 2.**

**Table 2:** Comparison between the nonlinear momentum model (NMM) and the experimental data.

| $I_p\ [W.cm^{-2}]$ | 2.4 | $6.6 \times 10^3$ | $5.5 \times 10^8$ | $3.7 \times 10^9$ |
|---|---|---|---|---|
| $N_{twin}^{exp}\ [Hz]$ | $1.1 \times 10^4$ | $4 \times 10^7$ | $3.7 \times 10^{14}$ | $1.2 \times 10^{21}$ |
| $N_{twin}^{NMM}\ [Hz]$ | $1.1 \times 10^3$ | $2.7 \times 10^6$ | $2.0 \times 10^{11}$ | $2.4 \times 10^{13}$ |

*3.1.1 The nonlinear Hamiltonian model*

The nonlinear Hamiltonian model (NHM) describes the quantum field in term of time dependent field mode operators assuming the UPA.[11] The calculation is detailed in **Appendix B**. Two intensity regimes can be also defined, one being linear, the other one exponential, which gives for the twin photon flux:

$$N_{twin}(Z) = \begin{cases} \dfrac{2\chi_{eff}^{(2)^2}}{\epsilon_0 n_s^2 n_i^2 n_p \lambda_s^2} \dfrac{n_{gs} n_{gi}}{\Delta n_g} \left|\dfrac{\sigma_p^2}{\sigma_s^2 + 2\sigma_p^2}\right|^2 I_p(0)Z, & \text{(small intensity)} \\ \dfrac{c}{2Ln_s} \exp(2\kappa_{time}Z), & \text{(high intensity),} \end{cases} \qquad (7)$$

with

$$\kappa_{time}^2 = \frac{\pi^2 {\chi_{eff}^{(2)}}^2}{4\epsilon_0 c n_s^2 \, n_i^2 n_p \lambda_p^2} \left|\frac{\sigma_p^2}{\sigma_s^2 + 2\sigma_p^2}\right|^2 I_p \tag{8}$$

The quantities $\sigma_{p,s}$ are the transverse section of both pump and down-converted fields, $n_{gs,i}$ are the group velocity indices of signal and idler, and $\Delta n_g = |n_{gs} - n_{gi}|$.

The curve corresponding to Equation 6 is shown in Figure 3 where it appears that it merges with the curve of the nonlinear momentum model. Some numerical values are given in **Table 3**.

**Table 3:** Comparison between the nonlinear Hamiltonian model (NHM) and the experimental data.

| $I_p \, [W \, cm^{-2}]$ | 2.4 | $6.6 \times 10^3$ | $5.5 \times 10^8$ | $3.7 \times 10^9$ |
|---|---|---|---|---|
| $N_{twin}^{exp} \, [Hz]$ | $1.1 \times 10^4$ | $4 \times 10^7$ | $3.7 \times 10^{14}$ | $1.2 \times 10^{21}$ |
| $N_{twin}^{NHM} \, [Hz]$ | $8.6 \times 10^2$ | $1.3 \times 10^6$ | $1.6 \times 10^{11}$ | $1.9 \times 10^{13}$ |

## 3.2 The semi-classical model

### 3.2.1 Coupled wave amplitudes equations

With the aim of improving the description of the experimental data of Figure 3, we proposed a semi-classical model, the so-called SCM, where the field motion is governed by the coupled wave amplitudes system coming from Maxwell's equations, the initial conditions being the pump field at the entrance of the medium and the quantum fluctuations of vacuum at the signal and idler wavelengths. The experiments were carried out in a phase-matched KTP crystal which is a non-conducting and non-magnetic medium. The involved wavelengths were in the transparency range of KTP, so that ABDP's assumption can be applied.[15] The propagation direction of the three interacting waves being along the x-axis of the dielectric frame, the spatial walk-off is then nil. There is no temporal walk-off either because the pump beams were emitted by CW, nanosecond or picosecond lasers. In these conditions, assuming the slowly-varying envelop, parallel beams and a collinear process, the coupled wave amplitudes system writes:

$$\begin{cases} \dfrac{\partial E_s(Z)}{\partial Z} = j\kappa_s \chi_{eff}^{(2)} E_p(Z) E_i^*(Z) \\ \dfrac{\partial E_i(Z)}{\partial Z} = j\kappa_i \chi_{eff}^{(2)} E_p(Z) E_s^*(Z) \\ \dfrac{\partial E_p(Z)}{\partial Z} = j\kappa_p \chi_{eff}^{(2)} E_s(Z) E_i(Z) \end{cases} \tag{9}$$

with

$$\kappa_a = \frac{\omega_a \epsilon_0 \mu_0 c}{2 n_a} \, (a \equiv p, s, i) \tag{10}$$

The spatial coordinate Z is along the direction of propagation, Z = 0 corresponding to the entrance of the crystal, $\chi_{eff}^{(2)}$ is the effective coefficient, $(E_p, E_s, E_i)$ are the complex amplitudes of the electric fields of the pump (p), signal (s) and idler (i) waves, $n_a (a \equiv p, s, i)$ is the refractive index at the circular frequency $\omega_a$.

The system of Equation 9 has analytical solutions that are Jacobi elliptic functions.[15] When the UPA is assumed, Equation 9 writes:

$$\begin{cases} \frac{\partial E_s(Z)}{\partial Z} = j\kappa_s \chi_{eff}^{(2)} E_p(0) E_i^*(Z) \\ \frac{\partial E_i(Z)}{\partial Z} = j\kappa_i \chi_{eff}^{(2)} E_p(0) E_s^*(Z) \\ \frac{\partial E_p(Z)}{\partial Z} = 0 \end{cases} \quad (11)$$

From Equation 11, it comes:

$$\begin{cases} \frac{\partial^2 E_i(Z)}{\partial Z} - \beta^2 E_i(Z) = 0 \\ E_s(Z) = j \frac{1}{\kappa_i \chi_{eff}^{(2)} E_p^*(0)} \frac{\partial E_i(Z)}{\partial Z} \end{cases} \quad (12)$$

where

$$\beta^2 = \chi_{eff}^{(2)} |E_p(0)| \sqrt{\kappa_s \kappa_i} \quad (13)$$

Equation 12 leads to the signal and idler motions described by hyperbolic functions, *i.e.*:

$$\begin{cases} E_s(Z) = E_s(0) ch(\beta Z) + j \sqrt{\frac{\omega_s n_i}{\omega_i n_s}} \frac{E_p(0)}{|E_p(0)|} E_i^*(0) sh(\beta Z) \\ E_i(Z) = E_i(0) ch(\beta Z) + j \sqrt{\frac{\omega_i n_s}{\omega_s n_i}} \frac{E_p(0)}{|E_p(0)|} E_s^*(0) sh(\beta Z) \end{cases} \quad (14)$$

The introduction of the quantum part of our semi-classical model is at the level of the initial conditions relative to the signal and idler amplitudes by setting:

$$\begin{cases} E_s(0) \equiv \Delta E_s^{vacuum} \\ E_i(0) \equiv \Delta E_i^{vacuum} \end{cases} \quad (15)$$

**Figure 4** gives a schematic physical view of the semi-classical model where it appears that the quantum fluctuations of vacuum stimulate the scission of the pump photons giving birth to signal and idler twin photons, so that the $sh(\beta Z)$ contributions describe the optical parametric amplification of vacuum while the $ch(\beta Z)$ contributions describe the generation over the mode not initially present.

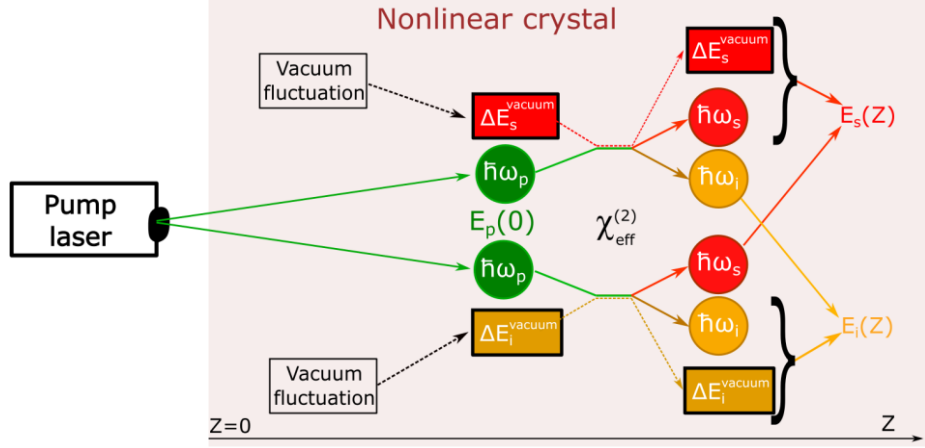

**Figure 4:** Schematic representation of the semi-classical modelling (SCM) of SPDC showing the initial fields $(E_p(Z=0), \Delta E_s^{vacuum}, \Delta E_i^{vacuum})$ and the generated fields $(E_s(Z), E_i(Z))$, where Z is the space coordinate along the direction of propagation. $\chi_{eff}^{(2)}$ is the second-order effective coefficient of the nonlinear crystal.

The calculation of the quantum fluctuations of vacuum is performed in the following section.

*3.2.2 Quantum fluctuations of the vacuum*

Like for the nonlinear momentum model, the electrical field quantum operator is expressed by use of continuous variable, considering a single spatial mode of propagation and taking the longitudinal cavity length as infinite.[10,16] The quantum field is modelled by space-dependent mode operators $a(\omega, Z)$ and writes:

$$E(t,Z) = i \int_0^{+\infty} d\omega \sqrt{\frac{\hbar\omega}{4\pi\epsilon_0 n(\omega)cS}} a(\omega, Z) e^{-i\omega(t-\frac{Z}{c})} + H.C. = E^+(t,Z) + E^-(t,Z) \quad (16)$$

Z is the longitudinal space coordinate, $n(\omega)$ is the refractive index of the nonlinear medium, and $S \approx \pi w_0^2$ the surface of the interaction volume where $w_0$ corresponds to the beam *radius*.

Quantum mechanically, the vacuum fluctuations are expressed using $\langle vac|\Delta^2 E(t,Z)|vac\rangle = \langle vac|E(t,Z)^2|vac\rangle = \langle vac|E^+(t,Z)E^-(t,Z) + E^-(t,Z)E^+(t,Z)|vac\rangle$. The corresponding calculation of the variance of the fluctuations $\Delta E^{vacuum}$ on a finite spectral range $\Delta\omega$ centered around a central frequency $\omega$ is detailed in **Appendix C**.[17] It leads to:

$$\Delta E^{vacuum} = \sqrt{\frac{\hbar\omega}{4\pi c\epsilon_0 n(\omega)S}\Delta\omega} \quad (17)$$

The spectral width $\Delta\omega$ can be estimated by the calculation of the SPDC spectral acceptance in the considered phase-matching direction, *i.e.* the x-axis of KTP, taken here the full-width at half-maximum (FWHM) of the phase-matching peak. An analytical expression can be obtained by approximating the phase-mismatch $\Delta k(\omega)$ by an affine function of slope $\alpha$ that cancels at $\omega_s$, *i.e.* the phase-matching circular frequency of the signal mode, which writes: $\Delta k(\omega) \approx \alpha(\omega - \omega_s)$. From Equation 2 using the dispersion equations of KTP, it comes $\alpha =$

$-1,97.10^{-9}\ m^{-1}Hz^{-1}$, as shown in **Figure 5** (Top).[13] Then the spectral width $\Delta\omega$ can be expressed as:

$$\Delta\omega \approx \frac{4}{|\alpha|L} \tag{18}$$

For $L = 1\ cm$, the spectral width is $\Delta\omega = 2.10^{11}\ Hz$ ; then taking $S = 2.5\times10^{-8}\ m^2$, $n(\omega) = 1.7$, it comes: $\Delta E_{s,i} \approx 12.5\ V.m^{-1}$. The vacuum fluctuation spectral density (VFSD) $\Delta^2 E_\omega$ is then integrated on the corresponding spectral interval $\Delta\omega$, leading to $\Delta^2 E$, as shown in Figure 5 (Down).

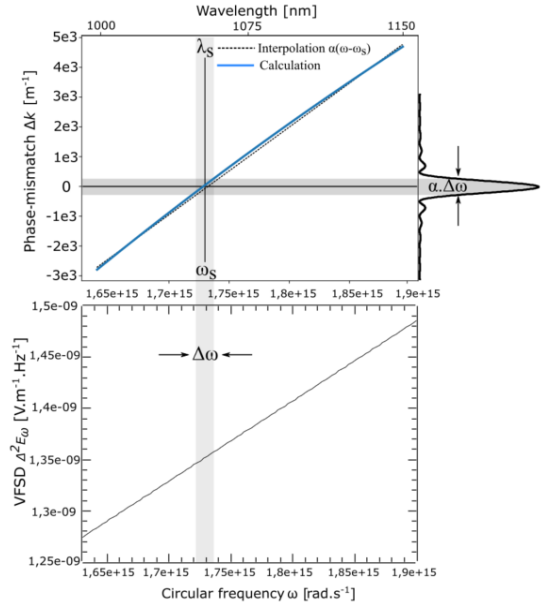

**Figure 5:** Schematic view of the quantum vacuum fluctuations calculation. (Top) The SPDC spectral acceptance is governed by the phase-mismatch $\Delta k(\omega)$, which defines a spectral width $\Delta\omega$. (Down) The quantity $\Delta^2 E_\omega$ is the vacuum fluctuation spectral density (VFSD).

*3.2.3 Twin photons flux*

Finally the twin photon flux $N_s = N_i = N_{twin}$ generated over the crystal length L and expressed in $[Hz]$ $(twins/s)$ can be deduced from the generated electrical field $E_s^{generated} = E_s - \Delta E_s^{vacuum}$ and using $N_s = \frac{\epsilon_0 n(\omega)cS}{4\hbar\omega}\left|E_s^{generated}\right|^2$. Then it comes:

$$N_{twin} = \frac{\epsilon_0 n_s cS}{4\hbar\omega_s}\left(|\Delta E_s^{vacuum}|(cosh(\beta L) - 1) + \sqrt{\frac{\omega_s n_s}{\omega_i n_i}}|\Delta E_i^{vacuum}|\sinh(\beta L)\right)^2 \tag{19}$$

Equation 19 is plotted Figure 3. It appears that the semi-classical model (SCM) is very close to the nonlinear momentum model (NMM) and to the nonlinear Hamiltonian model (NHM) in the linear part, contrary to the exponential part, *i.e.* above 30 MW.cm$^{-2}$, where the agreement of the SCM with the experimental data is much better than for the NMM and NHM. **Table 4** gives some numerical values that can directly be compared with those of Tables 2 and 3.

**Table 4:** Comparison between the semi-classical model (SCM) and the experimental data.

| $I_p \ [W \ cm^{-2}]$ | 2.4 | $6.6 \times 10^3$ | $5.5 \times 10^8$ | $3.7 \times 10^9$ |
|---|---|---|---|---|
| $N_{twin}^{exp}$ [Hz] | $1.1 \times 10^4$ | $4 \times 10^7$ | $3.7 \times 10^{14}$ | $1.2 \times 10^{21}$ |
| $N_{twin}^{BBB}$ [Hz] | $3.5 \times 10^2$ | $9.7 \times 10^5$ | $1.75 \times 10^{13}$ | $3.5 \times 10^{19}$ |

There is a discrepancy of one order of magnitude at 550 MW.cm$^{-2}$ for the SCM while it is seven order of magnitude in the case of the NMM (*cf* Table 2) and NHM (*cf* Table 3). And it is two orders of magnitude *versus* eight orders at 3.7 GW.cm$^{-2}$. The better agreement of the SCM at high intensity can be explained by simply considering that in this regime, the generated twin-photons amount is much bigger than the quantum fluctuations of vacuum. As a consequence, this range of intensity mainly corresponds to a coupling between classical fields, which is the framework of Equation 9, 10 and 11 coming from Maxwell's equations. In other words, the SCM becomes less and less of an approximation as pump intensity increases. It is exactly the reverse for the quantum models NMM and NHM that are a little bit better than the semi-classical model SCM at low pump intensity, as shown in Table 2, 3 and 4.

Note that the numerical integration of Equation 9, where there is no approximation regarding the pump depletion, leads to the same curve than that calculated with Equation 19 where the UPA is assumed, which is a validation of Equation 19.

The interpolation of the experimental data with the SCM leads to a full agreement in terms of magnitude and behaviour by taking $\Delta E_{s,i}^{vacuum} \approx 62.5 \ V.m^{-1}$, which is very close to the calculated value, *i.e.* $12.5 \ V.m^{-1}$. It is difficult at this step to find a satisfying explanation for this small gap, but it could provide from the fact that our modelling is performed under the collinear approximation while non-collinear contributions have been identified, as seen in Figure 2. The quantum models NMM and NHM can also fit well the experimental data with a multiplicative constant $\gamma = 3$ of the effective coefficient $\chi_{eff}^{(2)}$, with the same difficulty to interpret this result.

As already mentioned, the curves of Figure 3 exhibit a linear part and an exponential part, which can be explained by the existence of two different regimes of SPDC according to the relative value of $E_{s,i}$ and $\Delta E_{s,i}^{vacuum}$ : one below 30 MW/cm$^2$, defining the low pump intensity regime, for which $E_{s,i} < \Delta E_{s,i}^{vacuum}$, and the other one above 30 MW/cm$^2$, *i.e.* the high pump intensity regime, that corresponds to $E_{s,i} > \Delta E_{s,i}^{vacuum}$. From Equation 19, it is possible to obtain an analytical expression describing these two regimes. For that, $\frac{\omega_s n_s}{\omega_i n_i} \approx 1$ and $|\Delta E_s^{vacuum}| \approx |\Delta E_i^{vacuum}|$ are assumed, and the product $\beta L$ is taken as the relevant parameter for distinguishing the two regimes since this quantity is the argument of the *ch* and *sh* functions of Equation (19). Then it comes:

$$\begin{cases} \lim_{\beta L \ll 1} N_{twin} = \frac{\pi \chi_{eff}^{(2)^2}}{2\epsilon_0 c n_s n_i n_p \lambda_s^2 |a|} I_p L \\ \lim_{\beta L \gg 1} N_{twin} = \frac{1}{4\pi |a| L} exp(2\beta L) \end{cases} \quad (20)$$

Then $\beta L = 1$ allows us to define the pump intensity $I_p^{limit}(0)$ that delimits the two regimes. According to Equation 13, it writes:

$$I_p^{lim}(0) = \frac{n_p}{2\mu_0 c L^2 \left(\chi_{eff}^{(2)}\right)^2 \kappa_i \kappa_s} = 30 \; MW \; cm^{-2} \tag{21}$$

These 30 $MW \; cm^{-2}$ correspond exactly to the limit found on the experimental curve of Figure 3. It appears also that the NLM / NHM curve separates from the SCM curve from the same pump intensity.

## 4. Conclusion

We performed a second-order spontaneous parametric down-conversion experiment in a Type II phase-matched KTP crystal pumped at $\lambda_p = 532 \; nm$ and giving birth to a signal beam at $\lambda_s = 1037 \; nm$ and an idler at $\lambda_i = 1092 \; nm$ along the x-axis of the crystal. Three pump sources have been used: in the CW, nanosecond and picosecond regimes. That allowed us to cover a pump intensity range from 2.4 W.cm$^{-2}$ to 3.7 GW.cm$^{-2}$ leading to the generation of twin photons with a flux $N_{twin}$ from 1.1x10$^4$ Hz to 1.2x10$^{21}$ Hz, respectively, the corresponding quantum efficiencies $N_{twin} / N_p$ being 1.8x10$^{-11}$ and 9.2x10$^{-4}$. We identified two SPDC regimes regarding the pump intensity: one below 30 MW.cm$^{-2}$ for which the flux increases linearly, and the other one above 30 MW.cm$^{-2}$ for which it behaves exponentially. We proposed a semi-classical model under the undepleted pump approximation based on the quantum fluctuations of vacuum for the seeding of the process and a propagation ensured by classical fields, which allowed a satisfying description of these two regimes, with a much better agreement at high intensity than the quantum models based on the nonlinear momentum or the nonlinear Hamiltonian operators.[10, 11] However, the semi-classical and quantum approaches are complementary since the quantum models give access to a differentiation of these two regimes from the quantum point of view, *i.e.* a superposition of the vacuum state and the single biphoton Fock state at low pump intensity while multi-biphoton Fock states are involved at high pump intensity.[11] The following of this work will be quantum measurements at low and high pump intensities, as well as the use of our semi-classical model for the twin-photon generation by four-wave-mixing, and also for triple-photon generation.[18, 19]

**Appendix A: The nonlinear momentum model**

The starting point of the nonlinear momentum model (NMM), is the modelling of the electrical field operator as a single collinearly propagating spatial mode, described with space-dependent spectral-mode operators and taking the longitudinal periodicity as infinite, as done with Equation 16.[10] The second-order nonlinear momentum operator writes:

$$G_{nl}^{(2)}(Z) = \int_0^{+\infty} d\omega_p \int_0^{+\infty} d\omega \; \hbar\beta(\omega) \left(a_s(\omega, Z) \; a_i(\omega_p - \omega, Z) \; a_p^t(\omega_p, Z) e^{-i\Delta k(\omega)Z} + H.C.\right) \tag{A1}$$

where $\beta(\omega)$ is not the same quantity than that of Equation 13.[10] The equation of motion is then the following:

$$\frac{\partial a_{s,i}(\omega,Z)}{\partial Z} = -\frac{i}{\hbar}[a_{s,i}(\omega,Z), G_{nl}^{(2)}(Z)] \tag{A2}$$

Following Dayan's development leads to the expression of the photon flux spectral density $N_s(\omega,Z)$ given by Equation 5.[10] The terms in Equations 3 and 4 are rearranged with the most common quantities: the intensity $I_p(0)$, and the second-order effective coefficient $\chi_{eff}^{(2)}$. The relations between these quantities and those from Dayan are the following $\kappa = \sqrt{|C^{(2)}|}$, $I_p(0) = \hbar\omega_p \frac{N_p(0)}{S}$ and $\frac{N_p(0)}{S}(\beta(\omega)\sqrt{S})^2 = f^{(2)}(\omega)^2 I_p(0)\chi_{eff}^{(2)}{}^2$ where $N_p(0)$ is the pump photon flux expressed in $Hz$, and corresponds to the quantity denoted $I_p$ by Dayan.[10]

The photon flux spectral density of the signal and idler modes is given by $N_i(\omega,Z) = N_s(\omega_p - \omega, Z)$. The total twin photon flux is obtained by performing a numerical integration of the photon flux spectral density over the whole spectrum, i.e. $N_{twin}(Z) = N_s(Z) = N_i(Z) = \int_0^{+\infty} d\omega\, N_s(\omega,Z)$. There is no simple analytic expression of $N_{twin}(Z)$ in the general case.

For high intensities, a finite spectral bandwidth $\Delta_\omega^{strong}$ can be defined for the spectral components fulfilling $2\pi I_p(0)\chi^{(2)^2}f^{(2)}(\omega)^2 \gg \frac{\Delta k(\omega)^2}{4}$. It writes:

$$\Delta_\omega^{strong} = f^{(2)}\chi_{eff}^{(2)}\frac{\sqrt{8\pi I_p(0)}}{\alpha} \tag{A3}$$

For the spectral components within this bandwidth $\Delta_\omega^{strong}$, the photon flux spectral density is expressed as:

$$\lim_{\sqrt{|C^{(2)}(\omega)|}Z \gg 1} N_s(\omega,Z) \approx \frac{1}{2\pi}\exp\left(f^{(2)}\chi_{eff}^{(2)}\sqrt{8\pi I_p(0)}\, Z\right) \tag{A4}$$

For low pump intensities $\Delta_\omega^{strong} \approx 0$ and the pump intensity dependency is linear, i.e.:

$$\lim_{\sqrt{|C^{(2)}|}Z \ll 1} N_{twin}(Z) \approx \frac{\pi^2 \chi_{eff}^{(2)}{}^2}{2\epsilon_0 c n_p(\omega_p) n_s(\omega_1) n_i(\omega_p - \omega_1)\lambda_s^2|\alpha|} I_p Z \tag{A5}$$

The expression in the low intensity regime is very close to our semi-classical model (SCM) and differs only by a factor of $\pi$. Concerning the high intensity regime, this is hard to analytically compare the expressions because the NMM model does not provide an analytic expression of the twin photon flux $N_{twin}(Z)$. The numerical comparison is furnished on Figure 3 or table 2 and 4.

## Appendix B: The nonlinear Hamiltonian model

The nonlinear Hamiltonian model starts on a description of the displacement field operator $D(\vec{r},t)$ by use of time dependent mode operators $a^t_{\vec{k},s}(t)$ where $\vec{k}$ is the wave-vector and s the polarization state, those two parameters defining a given mode of propagation.[11] A basis change is operated to express those mode operators in a basis of transverse modes $a_{\vec{\mu},k_z,s}(t)$, where $\vec{\mu}$ is a pair of positive integer referring to the horizontal and vertical wave-vector components indexation, and $k_z$ is the longitudinal wave-vector component.[11] Those field modes are then expressed with Hermite-Gaussian wavefunction $g_{\vec{\mu}}(x,y)$. Assuming paraxial approximation, the displacement field writes:[11]

$$D(\vec{r},t) = i \sum_{\vec{\mu},k_z,s} \sqrt{\frac{\epsilon_0 n^2_{k_z} \hbar \omega_{k_z}}{2 L_z}} \vec{\epsilon_{k_z}} g_{\vec{\mu}}(x,y) e^{i k_z z} e^{-i\omega t} a^t_{\vec{\mu},k_z,s} + H.C. \tag{B1}$$

The nonlinear Hamiltonian is then expressed as follows:

$$H_{nl} = \frac{\hbar |E_p(0)|}{2 L_z} \sum_{\mu_s,k_{sz}} \sum_{\mu_s,k_{sz}} \sqrt{\frac{\omega_s \omega_i}{n_s^2 n_i^2}} \\ \times \int d^3 r \left( \chi^{(2)}_{eff}(\vec{r}) G^*_p(\vec{r}) g_{\overline{\mu_s}}(x,y) g_{\overline{\mu_i}}(x,y) e^{-i\Delta k_z} \right) e^{i\Delta\omega t} a^t_{\overrightarrow{\mu_s},k_{sz}} a^t_{\overrightarrow{\mu_i},k_{i_z}} \\ + H.C. \tag{B2}$$

The transition probability from the quantum state $|\psi(t)\rangle$ to a biphoton state $|\overrightarrow{\mu_s} k_{s_z}, \overrightarrow{\mu_i} k_{i_z}\rangle$ can be compute assuming a first-order approximation $P = |\langle \overrightarrow{\mu_s} k_{s_z}, \overrightarrow{\mu_i} k_{i_z} | \psi(t) \rangle|^2 \approx \left| \langle \overrightarrow{\mu_s} k_{s_z}, \overrightarrow{\mu_i} k_{i_z}| \left(1 - \frac{i}{\hbar} \int_0^t dt' H_{nl}(t')\right) 0,0 \rangle \right|^2$. From this expression derives the transition rate which in the case of a type II SPDC performed from a single mode undepleted pump beam leads to the expression given Equation 7. [11]

A more general expression can be obtained through the resolution of the Heisenberg's equation of motion.[11]

$$\frac{d a_{k_{s,i}}}{dt} = -\frac{i}{\hbar} [a_{k_{s,i}}, H_{nl}] \tag{B3}$$

This solving leads to the expression of the photon flux $N_s^{crystal}(t) \approx \sinh^2\left(\sqrt{N_p^{(0)}} |g| t\right)$ where $N_s^{crystal}(t)$ is the number of signal photons and $N_p^{(0)}$ the number of pump photons contained in the crystal at a time "t". The pump intensity $I_p(0)$ is linked to $N_p^{(0)}$ by $N_p^0 = \frac{I_p S}{2} \frac{\lambda_p}{hc} \frac{n_p L}{c}$. The twin photon flux at the output of the crystal $N_s(L)$ is linked to the number of twins in the crystal $N_s^{crystal}(t)$ by $N_s(L) = \frac{c}{L n_s} N_s^{crystal}(t = L/c)$. Finally, we obtain the second part of Equation 7.

**Appendix C: Quantum fluctuations of vacuum**

From Equation 16, the calculation of the quantum fluctuations of the vacuum electrical field is performed over a spectral range $\Delta\omega = \omega_{max} - \omega_{min}$ centred around $\omega$, forgetting about the terms $e^{\pm i\omega(t-\frac{Z}{c})}$ that cancels in the calculation. Then it comes:

$$\langle vac|\Delta^2 E(t,Z)|vac\rangle = \qquad (C1)$$

$$= \left\langle vac \left| \left( i \int_{\omega_{mini}}^{\omega_{maxi}} d\omega \sqrt{\frac{\hbar\omega}{4\pi c\epsilon_0 n(\omega)S}} a(\omega) \right) \left( -i \int_{\omega_{mini}}^{\omega_{maxi}} d\omega \sqrt{\frac{\hbar\omega}{4\pi c\epsilon_0 n(\omega)S}} a^t(\omega) \right) \right.\right.$$

$$+ \left. \left( -i \int_{\omega_{mini}}^{\omega_{maxi}} d\omega \sqrt{\frac{\hbar\omega}{4\pi c\epsilon_0 n(\omega)S}} a^t(\omega) \right) \left( i \int_{\omega_{mini}}^{\omega_{maxi}} d\omega \sqrt{\frac{\hbar\omega}{4\pi c\epsilon_0 n(\omega)S}} a(\omega) \right) \right| vac \right\rangle$$

$$= \left\langle vac \left| \left( \int_{\omega_{mini}}^{\omega_{maxi}} d\omega \frac{\hbar\omega}{4\pi c\epsilon_0 n(\omega)S} a(\omega) a^t(\omega) \right) \right| vac \right\rangle$$

$$= \left\langle vac \left| \left( \int_{\omega_{mini}}^{\omega_{maxi}} d\omega \frac{\hbar\omega}{4\pi c\epsilon_0 n(\omega)S} \left(1 + a^t(\omega)a(\omega)\right) \right) \right| vac \right\rangle$$

$$= \int_{\omega_{mini}}^{\omega_{maxi}} d\omega \frac{\hbar\omega}{4\pi c\epsilon_0 n(\omega)S} = \frac{\hbar}{8\pi c\epsilon_0 n(\omega)S}(\omega_{maxi}^2 - \omega_{mini}^2) \approx \frac{\hbar\omega}{4\pi c\epsilon_0 n(\omega)S}\Delta\omega$$

Then $\Delta E = \sqrt{\Delta^2 E}$ leads to Equation 17. This expression of $\Delta E$ is the same as in Riek *et al.* work but expressed in term of circular frequency $\omega$ rather than in frequency $\nu$, the comparison of this theoretical expression being in excellent agreement with the experimental data.[17]

The spectral door width is estimated by $sinc\left(\frac{\Delta k(\omega)L}{2}\right) = 0.5$, which leads to $\frac{\alpha\Delta\omega L}{2} \approx 2$ and finally Equation 18.

Acknowledgements

The authors wish to thank David Jegouso and Corinne Félix for their technical support concerning the laser sources. This work was funded in part by Fédération QuantAlps.

The table of contents entry should be 50–60 words long and should be written in the present tense. The text should be different from the abstract text.

C. Author 2, D. E. F. Author 3, A. B. Corresponding Author* ((same order as byline))

**Title** ((no stars))

ToC figure ((Please choose one size: 55 mm broad × 50 mm high **or** 110 mm broad × 20 mm high. Please do not use any other dimensions))

((Supporting Information can be included here using this template))

# Supporting Information

**Title** ((no stars))

*Author(s), and Corresponding Author(s)\** ((write out full first and last names))

((Please insert your Supporting Information text/figures here. Please note: Supporting Display items, should be referred to as Figure S1, Equation S2, etc., in the main text…)